\documentclass[
  aps,
  prapplied,
  twocolumn,
  reprint,
  superscriptaddress,
  floatfix,
  citeautoscript,
  longbibliography,
]{revtex4-2}

\usepackage[T1]{fontenc}
\usepackage[utf8]{inputenc}

\usepackage{newtxtext}
\usepackage{newtxmath}

\usepackage{array}
\newcolumntype{C}[1]{>{\centering}m{#1}}

\usepackage{chemformula}
\usepackage{siunitx}
\DeclareSIUnit\atom{atom}
\DeclareSIUnit\gauss{G} 
\DeclareSIUnit\torr{Torr}
\DeclareSIUnit{\wtpercent}{wt.~\%}

\usepackage{graphicx}
\graphicspath{{./figures-3}}

\usepackage[
   unicode,
   colorlinks = true,
   allcolors = blue,
]{hyperref}
\usepackage{booktabs,multirow}

\newcommand*{\figref}[2][]{%
  \hyperref[{fig:#2}]{%
    Figure~\ref*{fig:#2}%
    \ifx\\#1\\%
    \else
      \,#1%
    \fi
  }%
}

\usepackage{cleveref}

\usepackage{microtype}

\usepackage{comment}
\usepackage[caption=false]{subfig} 
\usepackage[normalem]{ulem}

\usepackage{tikz}
\PassOptionsToPackage{usenames,dvipsnames}{xcolor}

\usepackage{glossaries}
\glsdisablehyper
\newacronym{1d}{1D}{one-dimensional}
\newacronym{2d}{2D}{two-dimensional}
\newacronym{3d}{3D}{three-dimensional}

\newacronym{ac}{AC}{alternating current}
\newacronym{afm}{AFM}{atomic force microscopy}
\newacronym{alc}{ALC}{avoided level crossing}
\newacronym{api}{API}{application programming interface}
\newacronym{ariel}{ARIEL}{Advanced Rare Isotope Laboratory}
\newacronym{arpes}{ARPES}{angle-resolved photoemission spectroscopy}
\newacronym{atp}{ATP}{adenosine triphosphate}
\newacronym{accg}{\ensuremath{E_{\mathrm{acc}}}}{accelerating gradient}
\newacronym{H0}{\ensuremath{H_{0}}}{applied field}

\newacronym[sort={b-NMR}]{bnmr}{\ensuremath{\beta}-NMR}{\ensuremath{\beta}-detected nuclear magnetic resonance}
\newacronym[sort={b-NQR}]{bnqr}{\ensuremath{\beta}-NQR}{\ensuremath{\beta}-detected nuclear quadrupole resonance}
\newacronym{bca}{BCA}{binary collision approximation}
\newacronym{bcc}{BCC}{body-centred cubic}
\newacronym{bcp}{BCP}{buffered chemical polishing}
\newacronym{bcs}{BCS}{Bardeen-Cooper-Schrieffer}
\newacronym{bl}{BL}{Bean-Livingston}
\newacronym{bpp}{BPP}{Bloembergen-Purcell-Pound}
\newacronym{bsc}{BSC}{\ch{Bi2Se3:Ca}}
\newacronym{btm}{BTM}{\ch{Bi2Te3:Mn}}
\newacronym{bts}{BTS}{\ch{Bi2Te2Se}}

\newacronym{camp}{CAMP}{control and monitor program}
\newacronym{ccd}{CCD}{charge-coupled device}
\newacronym{cdw}{CDW}{charge density wave}
\newacronym{cgs}{CGS}{centimetre-gram-second system of units}
\newacronym{cmms}{CMMS}{Centre for Molecular and Materials Science}
\newacronym{codata}{CODATA}{Committee on Data for Science and Technology}
\newacronym{cpu}{CPU}{central processing unit}
\newacronym{create}{CREATE}{Collaborative Research and Training Experience Program}
\newacronym{cw}{CW}{continuous wave}

\newacronym{daq}{DAQ}{data acquisition}
\newacronym{dc}{DC}{direct current}
\newacronym{dft}{DFT}{density functional theory}
\newacronym{dos}{DOS}{density of states}
\newacronym{dqt}{DQT}{double-quantum transition}

\newacronym{efg}{EFG}{electric field gradient}
\newacronym{emim-ac}{EMIM-Ac}{1-ethyl-3-methylimidazolium acetate}
\newacronym{emim-dca}{EMIM-DCA}{1-ethyl-3-methylimidazolium dicyanamide}
\newacronym{epr}{EPR}{electron paramagnetic resonance}
\newacronym{esr}{EPR}{electron spin resonance}
\newacronym{endor}{ENDOR}{electron nuclear double resonance}
\newacronym{epics}{EPICS}{Experimental Physics and Industrial Control System}
\newacronym{ep}{EP}{electropolishing}
\newacronym{edx}{EDX}{energy dispersive X-ray spectroscopy}

\newacronym{fcc}{FCC}{face-centred cubic}
\newacronym{fft}{FFT}{fast Fourier transform}
\newacronym{fom}{FoM}{figure of merit}
\newacronym{fwhm}{FWHM}{full width at half maximum}
\newacronym{ffvp}{\ensuremath{B_{\mathrm{vp}}}}{field of first vortex penetration}

\newacronym{gga}{GGA}{generalized gradient approximation}
\newacronym{gl}{GL}{Ginzburg-Landau}

\newacronym{hb}{HB}{hole-burning}
\newacronym{hfqs}{HFQS}{high-field \ensuremath{Q} slope}
\newacronym{hv}{HV}{high-voltage}
\newacronym{hwhm}{HWHM}{half width at half maximum}
\newacronym{hpcvd}{HPCVD}{hybrid physical chemical vapour deposition}

\newacronym{iaea}{IAEA}{International Atomic Energy Agency}
\newacronym{il}{IL}{ionic liquid}
\newacronym{is}{IS}{impedance spectroscopy}
\newacronym{isac}{ISAC}{isotope separator and accelerator}
\newacronym{isol}{ISOL}{isotope separation online}
\newacronym{isosim}{IsoSiM}{Isotopes for Science and Medicine}

\newacronym{jlab}{JLab}{Thomas Jefferson National Accelerator Facility}

\newacronym{lcao}{LCAO}{linear combination of atomic orbitals}
\newacronym{lda}{LDA}{local density approximation}
\newacronym{led}{LED}{light-emitting diode}
\newacronym{leis}{LEIS}{low-energy ion scattering}
\newacronym{lib}{LIB}{lithium-ion battery}
\newacronym{lsat}{LSAT}{\ch{(La,Sr)(Al,Ta)O3}}
\newacronym{ltb}{LTB}{low-temperature baking}

\newacronym{mas}{MAS}{magic angle spinning}
\newacronym{mpms}{MPMS}{magnetic property measurement system}
\newacronym{mbe}{MBE}{molecular beam epitaxy}
\newacronym{md}{MD}{molecular dynamics}
\newacronym{midas}{MIDAS}{Maximum Integrated Data Acquisition System}
\newacronym{mit}{MIT}{metal-insulator transition}
\newacronym{mnr}{MNR}{Meyer-Neldel rule}
\newacronym{mqt}{mqt}{multi-quantum transition}
\newacronym{mud}{MUD}{muon data}
\newacronym{ms}{MS}{mass spectrometry}
\newacronym{bmax}{\ensuremath{B_\mathrm{max}}}{maximum field in superconducting
heterostructures that can be sustained while remaining in the
Meissner state}

\newacronym{nbm}{NBM}{neutral beam monitor}
\newacronym{neb}{NEB}{nudged elastic band}
\newacronym{nim}{NIM}{nuclear instrumentation module}
\newacronym{nmr}{NMR}{nuclear magnetic resonance}
\newacronym{no}{NO}{nuclear orientation}
\newacronym{nqr}{NQR}{nuclear quadrupole resonance}
\newacronym{nrc}{NRC}{National Research Council of Canada}
\newacronym{nserc}{NSERC}{Natural Sciences and Engineering Research Council of Canada}

\newacronym{oa}{OA}{optical absorption}

\newacronym{pac}{PAC}{perturbed angular correlation}
\newacronym{pad}{PAD}{perturbed angular distribution}
\newacronym{pas}{PAS}{principle axis system}
\newacronym{pchip}{PCHIP}{piecewise cubic Hermite interpolating polynomial}
\newacronym{pdf}{PDF}{probability density function}
\newacronym{pld}{PLD}{pulsed laser deposition}
\newacronym{ppms}{PPMS}{physical property measurement system}
\newacronym{psi}{PSI}{Paul Scherrer Institute}

\newacronym{qens}{QENS}{quasielastic neutron scattering}
\newacronym{ql}{QL}{quintuple layer}
\newacronym{qo}{QO}{quantum oscillations}
\newacronym{qf}{\ensuremath{Q}}{quality factor}

\newacronym{rbs}{RBS}{Rutherford backscattering}
\newacronym{rf}{RF}{radio frequency}
\newacronym{rheed}{RHEED}{reflection high-energy electron diffraction}
\newacronym{rib}{RIB}{radioactive ion beam}
\newacronym{rkky}{RKKY}{Ruderman–Kittel–Kasuya–Yosida}
\newacronym{rrr}{RRR}{residual-resistivity ratio}
\newacronym{rtil}{RTIL}{room temperature ionic liquid}

\newacronym{sae}{SAE}{spin-alignment echo}
\newacronym{sans}{SANS}{small angle neutron scattering}
\newacronym{si}{SI}{International System of Units}
\newacronym{sis}{SIS}{superconductor-insulator-superconductor}
\newacronym{sims}{SIMS}{secondary ion mass spectrometry}
\newacronym{slr}{SLR}{spin-lattice relaxation}
\newacronym[sort={S/N}]{snr}{\textit{S}/\textit{N}}{signal-to-noise ratio}
\newacronym{squid}{SQUID}{superconducting quantum interference device}
\newacronym{srf}{SRF}{superconducting radio frequency}
\newacronym{srim}{SRIM}{Stopping and Range of Ions in Matter}
\newacronym{ss}{SS}{superconductor-superconductor}
\newacronym{ssid}{SSID}{solid-state ionic device}
\newacronym{ssr}{SSR}{spin-spin relaxation}
\newacronym{stm}{STM}{scanning tunnelling microscopy}
\newacronym{sts}{STS}{scanning tunnelling spectroscopy}
\newacronym{sh}{\ensuremath{H_\mathrm{sh}}}{superheating field}
\newacronym{sem}{SEM}{scanning electron microscope}

\newacronym{ti}{TI}{topological insulator}
\newacronym{tem}{TEM}{transmission electron microscopy}
\newacronym{trim}{TRIM}{Transport and Range of Ions in Matter}
\newacronym{tss}{TSS}{topological surface state}
\newacronym{tmd}{TMD}{transition metal dichalcogenide}
\newacronym{tofsims}{TOF-SIMS}{time-of-flight secondary ion mass spectrometry}

\newacronym{uhv}{UHV}{ultra-high vacuum}

\newacronym{vdw}{vdW}{van der Waals}
\newacronym{vft}{VFT}{Vogel-Fulcher-Tammann}

\newacronym{xrd}{XRD}{x-ray diffraction}
\newacronym{xrr}{XRR}{x-ray reflection}

\newacronym{ybco}{YBCO}{\ch{YBa2Cu3O_{6+x}}}
\newacronym{ysz}{YSZ}{yttria-stabilized zirconia}

\newacronym[sort={muSR}]{musr}{\ensuremath{\mu}SR}{muon spin rotation}
\newacronym{alc-musr}{ALC-\ensuremath{\mu}SR}{avoided level crossing muon spin rotation}
\newacronym{le-musr}{LE-\ensuremath{\mu}SR}{low-energy muon spin rotation}
\newacronym{lf-musr}{LF-\ensuremath{\mu}SR}{longitudinal field muon spin rotation}
\newacronym{rf-musr}{RF-\ensuremath{\mu}SR}{radio frequency muon spin rotation}
\newacronym{tf-musr}{TF-\ensuremath{\mu}SR}{transverse field muon spin rotation}
\newacronym{zf-musr}{ZF-\ensuremath{\mu}SR}{zero field muon spin rotation}

\begin{document}

\title{
  Measurements of the first-flux-penetration field in surface-treated and coated Nb: \\
  Distinguishing between near-surface pinning and an interface energy barrier
}

\author{Md~Asaduzzaman}
\email[E-mail: ]{asadm@uvic.ca}
\affiliation{Department of Physics and Astronomy, University of Victoria, 3800 Finnerty Road, Victoria, BC V8P~5C2, Canada}
\affiliation{TRIUMF, 4004 Wesbrook Mall, Vancouver, BC V6T~2A3, Canada}

\author{Ryan~M.~L.~McFadden}
\affiliation{Department of Physics and Astronomy, University of Victoria, 3800 Finnerty Road, Victoria, BC V8P~5C2, Canada}
\affiliation{TRIUMF, 4004 Wesbrook Mall, Vancouver, BC V6T~2A3, Canada}

\author{Edward~Thoeng}
\affiliation{TRIUMF, 4004 Wesbrook Mall, Vancouver, BC V6T~2A3, Canada}
\affiliation{Department of Physics and Astronomy, University of British Columbia, 6224 Agricultural Road, Vancouver, British Columbia V6T 1Z1, Canada}

\author{Robert~E.~Laxdal}
\affiliation{TRIUMF, 4004 Wesbrook Mall, Vancouver, BC V6T~2A3, Canada}

\author{Tobias~Junginger}
\email[E-mail: ]{junginger@uvic.ca}
\affiliation{Department of Physics and Astronomy, University of Victoria, 3800 Finnerty Road, Victoria, BC V8P~5C2, Canada}
\affiliation{TRIUMF, 4004 Wesbrook Mall, Vancouver, BC V6T~2A3, Canada}

\date{\today}

\begin{abstract}
  We report measurements of the first-flux-penetration field in surface-treated
  and coated Nb samples using \gls{musr}.
  Using thin \ch{Ag} foils as energy moderators for the implanted muon spin-probes,
  we ``profile'' the vortex penetration field $\mu_{0} H_{\mathrm{vp}}$ at sub-surface depths
  on the order of \qtyrange{\sim 10}{\sim 100}{\micro\meter}.
  In a coated sample [Nb$_3$Sn(\qty{2}{\micro\meter})/Nb],
  we find that $\mu_{0} H_{\mathrm{vp}}$ is depth-independent with a value of
  \qty{234.5 \pm 3.5}{\milli\tesla},
  consistent with Nb's metastable superheating field
  and
  suggestive of surface energy barrier for flux penetration.
  Conversely,
  in a surface-treated sample [Nb baked in vacuum at \qty{120}{\celsius} for \qty{48}{\hour}],
  vortex penetration onsets close to pure Nb's lower critical field
  $\mu_{0}H_\mathrm{c1} \approx \qty{170}{\milli\tesla}$,
  but increases with increasing implantation depth,
  consistent with flux-pinning localized at the surface.
  The implication of these results for technical applications of superconducting
  Nb, such as \gls{srf} cavities,
  is discussed.
\end{abstract}

\maketitle
\glsresetall

\section{
  Introduction
  \label{sec:introduction}
 }

A key technical application of the elemental type-II superconductor Nb
is its use in \gls{srf}
cavities~\cite{2008-Padamsee-RFSA-2,2009-Padamsee-RFSSTA,2023-Padamsee-SRTA},
which are utilized in particle accelerators across the globe.
Crucial to their operation is maintaining Nb in its
magnetic-flux-free Meissner state
(i.e., to prevent dissipation caused by magnetic vortices),
which generally restricts their use to surface magnetic fields up to
the element's lower critical field
$\mu_{0} H_\mathrm{c1} \approx \qty{170}{\milli\tesla}$~\cite{2022-Turner-SR-12-5522}.
Such a limitation ultimately sets a ceiling for a cavity's maximum
accelerating gradient $E_{\mathrm{acc}}$
(i.e., the achievable energy gain per unit length),
which impacts design considerations for accelerating structures
(e.g., size, operating temperature, etc.).
Consequently,
there is great interest in pushing \gls{srf} cavity operation up to
Nb's so-called superheating field
$\mu_{0}H_{\mathrm{sh}} \approx \qty{240}{\milli\tesla}$~\cite{2015-Posen-PRL-115-047001,2017-Junginger-SST-30-125012},
where the Meissner state is preserved in a metastable configuration.
Currently,
the largest 
gradients are achieved by so-called \gls{ltb} surface treatments,
wherein a Nb cavity is baked at temperatures on the order of \qty{\sim 120}{\celsius}
either in
vacuum~\cite{2004-Ciovati-JAP-96-1591,arXiv:1806.09824}
or
in a low-pressure gas atmosphere~\cite{2013-Grassellino-SST-26-102001,2017-Grassellino-SST-30-094004}.
Indeed,
the best performing treatments have enabled cavities to achieve surface magnetic fields beyond $\mu_{0} H_\mathrm{c1}$
(with some even approaching $H_\mathrm{sh}$)~\cite{2015-Posen-PRL-115-047001};
however,
the underlying mechanism for this enhancement
remains unclear.

Consider the typical \gls{ltb} treatment,  involving
vacuum annealing Nb at \qty{120}{\celsius}
for a duration up to \qty{48}{\hour}~\cite{2004-Ciovati-JAP-96-1591}.
Early measurements of this treatment's effect on Nb's
Meissner response using
\gls{le-musr}~\cite{2004-Morenzoni-JPCM-16-S4583,2004-Bakule-CP-45-203}
showed a sharp discontinuity in the screening
profile~\cite{2014-Romanenko-APL-104-072601},
which led to suggestions that \gls{ltb} can be used to create
an ``effective'' \gls{ss} bilayer~\cite{2017-Kubo-SST-30-023001}.
Of particular interest for \gls{srf} applications is the case
where a thin ``dirty'' layer resides atop a ``clean'' bulk
(e.g., as a result of surface-localized inhomogeneous doping),
as it offered several avenues for increasing
the vortex penetration field $\mu_{0}H_{\mathrm{vp}}$
via a reduced current density at Nb's surface
(e.g.,
following directly from the \gls{ss}-like structure~\cite{2017-Kubo-SST-30-023001}
or
as a consequence of deformations found in the
Meissner profile itself~\cite{2019-Ngampruetikorn-PRR-1-012015,2020-Checchin-APL-117-032601}).
While both theories~\cite{2017-Kubo-SST-30-023001} and measurements~\cite{2014-Romanenko-APL-104-072601} have been presented that support the interpretation
of a surface barrier originating from a ``dirty'' layer,
other measurements~\cite{2023-McFadden-PRA-19-044018} and analyses~\cite{2024-Ryan-APL-124-086101} contradict such views.
To resolve this discrepancy,
additional measurements using alternative
approaches would be highly beneficial.

One such possibility is instead using techniques
capable of identifying $\mu_{0}H_{\mathrm{vp}}$ directly.
This has been done,
for example,
using ``bulk'' \gls{musr}~\cite{2011-Yaouanc-MSR,2022-Hillier-NRMP-2-4},
which provides a local measurement of the magnetic field
\qty{\sim 100}{\micro\meter} below Nb's surface.
Such studies have found $\mu_{0} H_{\mathrm{vp}} \gtrsim \mu_{0} H_\mathrm{c1}$
for both \gls{ltb}
and
coated Nb~\cite{2018-Junginger-PRAB-21-032002},
the latter yielding
$\mu_{0}H_{\mathrm{vp}} \approx \mu_{0}H_{\mathrm{sh}}$~\cite{2017-Junginger-SST-30-125012}.
While this provided strong evidence that a
surface energy barrier~\cite{1964-Bean-PRL-12-14}
was preventing flux nucleation in \gls{ss} samples,
some ambiguity in interpreting the enhancement from \gls{ltb} remained.
Specifically,
subsequent magnetometry measurements on identically prepared samples
showed no such enhancement~\cite{2022-Turner-SR-12-5522},
implying an accumulation of near-surface vortices caused by \emph{pinning}.
The discrepancy between \gls{musr} and magnetometry
suggests that any pinning centers must be localized to depths
less than \qty{\sim 100}{\micro\meter} below Nb's surface.

\begin{figure}
  \centering
  \subfloat[Meissner state]{\includegraphics[width=0.225\textwidth]{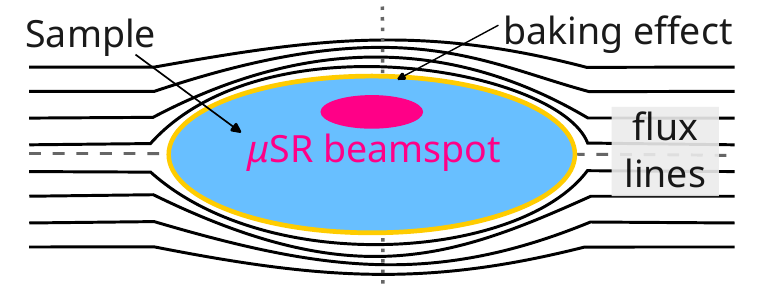}\label{fig:Meissner-state}}
  \subfloat[Vortex state]{\includegraphics[width=0.225\textwidth]{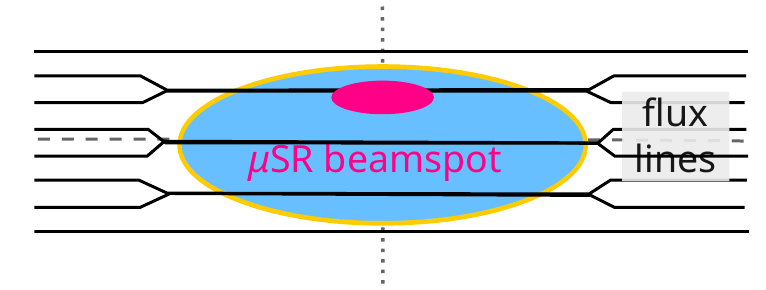}\label{fig:vortex-state}}
  \vskip\baselineskip
  \subfloat[Vortex state with (undetected) pinning]{\includegraphics[width=0.225\textwidth]{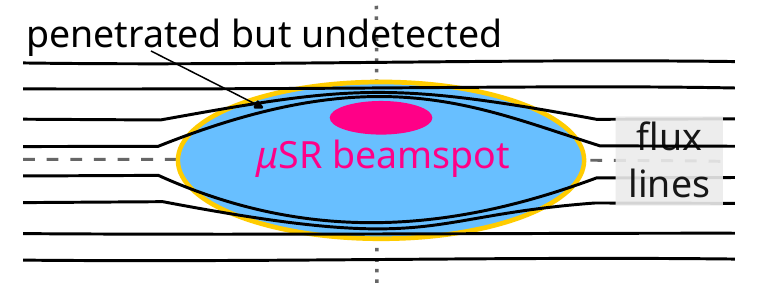}\label{fig:vs-pin-undetected}}
  \subfloat[Vortex state with (detected) pinning]{\includegraphics[width=0.225\textwidth]{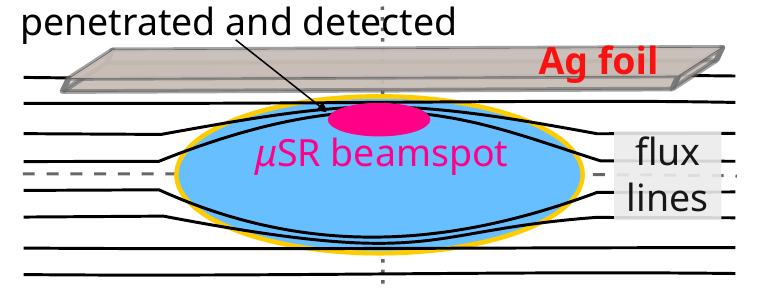}\label{fig:vs-pin-detected}}
  \caption{
    \label{fig:experiment-sketch}
    Sketch of the present \gls{musr} experiment on superconducting Nb samples
    with ellipsoidal shape in an applied magnetic field parallel to the ellipsoid
    major axis (- - -), with the magnetic flux lines (---) also indicated.
    (\textbf{a}):
    In the Meissner state, complete flux expulsion from the ellipsoid's interior
    is achieved.
    Without any energy moderation for the $\mu^{+}$ beam,
    the magnetic probes stop well-below the sample surface and experience no
    external contribution to their local field.
    (\textbf{b}):
    In the vortex state,
    some magnetic flux penetrates the sample as quantized fluxoids with a
    field-depended lattice arrangement, leading to a broad local field distribution
    samples by the $\mu^{+}$ beamspot.
    (\textbf{c}):
    In the presence of strong near-surface pinning in the vortex state,
    fluxoid penetration is localized to the sample surface,
    which may go undetected by the implanted $\mu^{+}$ at full beam energy.
    (\textbf{d}):
    Through the use of thin \ch{Ag} foils as energy moderators for the $\mu^{+}$ beam,
    the $\mu^{+}$ probes stop closer to the surface,
    allowing for flux that is surface-pinned in the vortex state to be observed.
  }
\end{figure}

To test these ideas,
here we extend the ``bulk'' \gls{musr} approach used in related
work~\cite{2013-Grassellino-PRSTAB-16-062002,2017-Junginger-SST-30-125012,2018-Junginger-PRAB-21-032002}
to provide \emph{depth-resolved} measurements
of $\mu_{0} H_{\mathrm{vp}}$
in both surface-treated and coated Nb.
Specifically,
we make use of thin \ch{Ag} foils to moderate the implantation energy of
the muon spin probes,
providing spatial sensitivity to depths on the
order of \qtyrange{\sim 30}{\sim 100}{\micro\meter}.
In the presence of surface energy barrier~\cite{1964-Bean-PRL-12-14},
$\mu_{0} H_{\mathrm{vp}}$ is expected to be depth-independent,
whereas surface-localized pinning is anticipated to produce
larger $\mu_{0} H_{\mathrm{vp}}$s deeper below the surface
(see \Cref{fig:experiment-sketch}).
Using this approach,
we find that
$\mu_{0} H_{\mathrm{vp}}$ is depth-independent
and
close to Nb's $\mu_{0} H_{\mathrm{sh}}$
for a Nb sample coated with the A15 superconductor
Nb$_3$Sn~\cite{2006-Godeke-SST-19-R68,2017-Posen-SST-30-033004},
consistent with the energy barrier expected for the \gls{ss} bilayer.
Conversely,
for Nb that has been surface-treated by \gls{ltb} at \qty{120}{\celsius},
the measured $\mu_{0} H_{\mathrm{vp}}$s are comparable to
Nb's $\mu_{0} H_\mathrm{c1}$,
but increase deeper below the surface,
suggesting the presence of
localized pinning near the surface that prevents detection by deeper implanted
muons.

\section{
  Experiment
  \label{sec:experiment}
 }

\begin{figure}
  \centering
  \includegraphics*[width=\columnwidth]{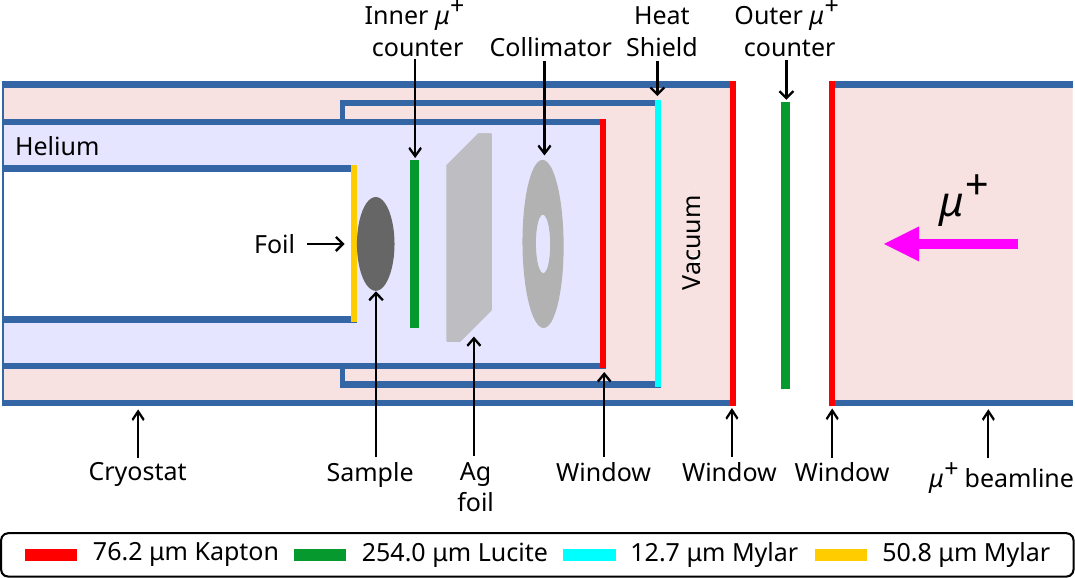}
  \caption{
    \label{fig:cryostat}
    Schematic of the horizontal gas-flow cryostat
    and
    low-background (i.e., Knight shift) insert
    used with TRIUMF's high parallel-field
    (i.e., ``HodgePodge'')
    spectrometer~\cite{2018-Kreitzman-JPSCP-21-011056}.
    The thin \ch{Ag} foils used as $\mu^{+}$ energy moderators
    are located between a \qty{8}{\milli\meter} diameter beam collimator
    (also made of \ch{Ag})
    and the inner $\mu^{+}$ counter.
  }
\end{figure}

\begin{figure}
  \centering
  \includegraphics*[width=\columnwidth]{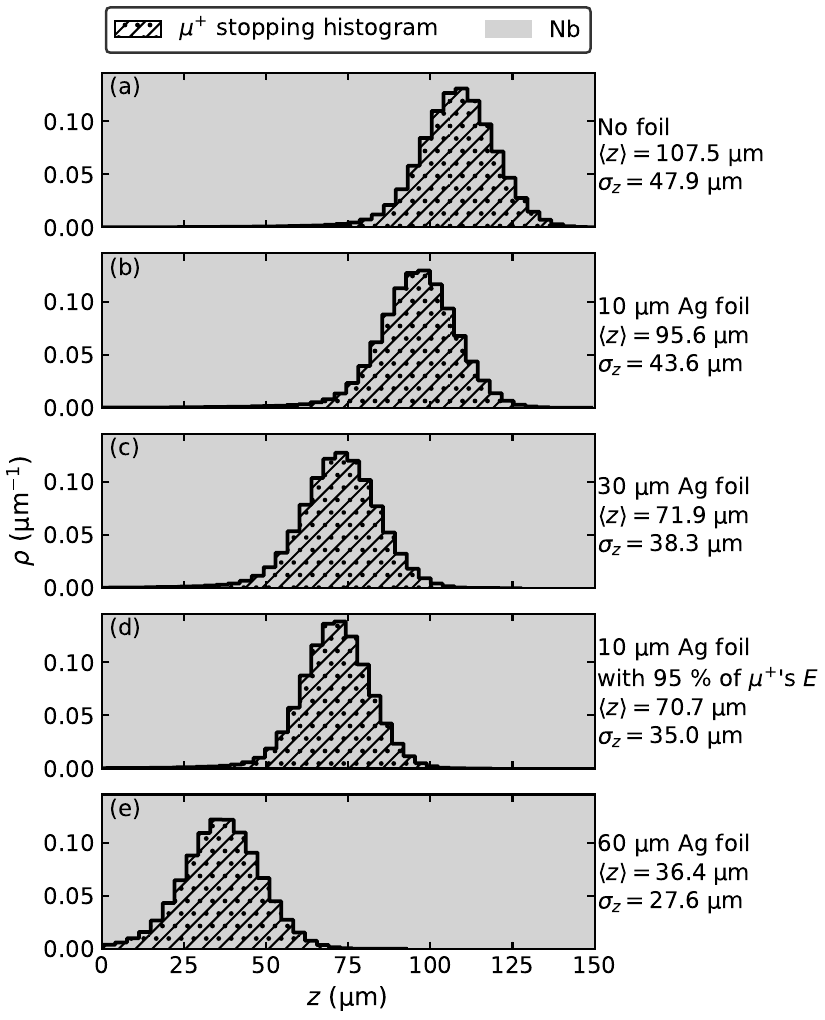}
  \caption{
    \label{fig:stopping-profiles}
    Simulated stopping profiles for \qty{\sim 4.1}{\mega\electronvolt}
    ``surface'' $\mu^{+}$ implanted in Nb using the
    \gls{srim} Monte Carlo code~\cite{srim}.
    The profiles,
    represented here as histograms,
    were generated from \num{e6} $\mu^{+}$ projectiles
    and
    account for all materials in the beam's path prior to
    implantation
    (e.g., cryostat windows, $\mu^{+}$ counters, moderating foils, etc. --- see \Cref{fig:cryostat}).
    Using \ch{Ag} foils of different thicknesses (indicated in each plot's inset), mean stopping depths $\langle z \rangle$ in the range of \qtyrange{\sim 36}{\sim 108}{\micro\meter} are achieved. The width (i.e., standard deviation) $\sigma_{z}$ of each stopping distribution is also indicated.
    Note that a reduced $\mu^{+}$ implantation energy is used for panel (d),
    yielding a $\langle z \rangle$ comparable to using a thicker moderating foil,
    as shown in panel (c).
  }
\end{figure}

\Gls{musr} experiments were performed at TRIUMF's \gls{cmms}
facility in Vancouver, Canada.
Using the M20C beamline~\cite{2018-Kreitzman-JPSCP-21-011056},
a \qty{\sim 100}{\percent} spin-polarized
\qty{\sim 4.1}{\mega\electronvolt} ``surface'' $\mu^{+}$ beam
was extracted,
spin-rotated in flight,
and
delivered to the high-parallel-field
(i.e., ``HodgePodge'')
spectrometer
equipped with a horizontal gas-flow cryostat
and a low-background
(i.e., Knight shift)
insert~\cite{2018-Kreitzman-JPSCP-21-011056}.
A sketch of cryostat configuration is given in \Cref{fig:cryostat}.
This setup is similar to that used in
related experiments~\cite{2013-Grassellino-PRSTAB-16-062002,2017-Junginger-SST-30-125012,2018-Junginger-PRAB-21-032002},
with the external magnetic field parallel to each sample's surface
(see \Cref{fig:experiment-sketch})
and perpendicular to the implanted $\mu^{+}$ spin direction.
Notably,
the present work also incorporates thin \ch{Ag} foils
(Goodfellow, \qty{99.95}{\percent} purity, \qtyrange{10}{30}{\micro\meter} thick)
as part of the cryostat assembly,
acting as energy moderators for the $\mu^{+}$ beam.
By varying the thickness of the foils,
the range of implanted $\mu^{+}$ in the Nb samples
can be controlled on the \unit{\micro\meter} scale. 
Simulations of the moderating effect were performed using the
\gls{srim} Monte Carlo code~\cite{srim},
which incorporated all major materials along the beam's path
(e.g., muon counters, moderator foils, etc. --- see \Cref{fig:cryostat}),
as well as compound corrections~\cite{1988-Ziegler-NIMB-35-215}
to the stopping powers (where appropriate).
Typical $\mu^{+}$ stopping profiles for this setup
are shown in \Cref{fig:stopping-profiles},
showing mean stopping ranges $\langle z \rangle$ between
\qtyrange{\sim 36}{\sim 108}{\micro\meter}
for a Nb target, along with the width (i.e., standard deviation) $\sigma_{z}$ of the stopping distributions~\footnote{Note that the mean stopping depth $\langle z \rangle$ is often used as a proxy for the full $\mu^{+}$ stopping distribution}.
Note that similar simulations for \ch{Nb_3Sn}(\qty{2}{\micro\meter})/Nb samples (not shown) yielded virtually identical results.
In cases where a \qty{60}{\micro\meter} \ch{Ag} foil was used, a small fraction
(\qty{\sim 2}{\percent})
of the implanted probes stop in the inner $\mu^{+}$ counter,
located immediately upstream of the sample
(see \Cref{fig:cryostat}).

In \gls{musr},
the implanted $\mu^{+}$ spin probes
(spin $S = 1/2$,
gyromagnetic ratio $\gamma / (2 \pi) = \qty{135.54}{\mega\hertz\per\tesla}$,
lifetime $\tau_{\mu} = \qty{2.197}{\micro\second}$)
are sensitive to the local magnetic field
at their stopping sites,
with the temporal evolution of the ensemble's spin-polarization
monitored via the anisotropic property of radioactive $\beta$-decay.
Specifically,
each $\mu^{+}$ decay positron's emission direction is
statistically correlated with the $\mu^{+}$ spin direction
at the moment of decay,
providing an easy means of tracking spin-reorientation.
Specifically,
in a two-detector setup
like that used here
(see \Cref{fig:cryostat}),
the recorded histogram of decay events takes the form:
\begin{equation}
  N_{\pm}(t) = N_{\pm, 0} \exp \left ( - \frac{t}{\tau_{\mu}} \right ) \left [ 1 \pm A_{0} P_{\mu}(t) \right ] + b_{\pm}
  \label{eq:counts}
\end{equation}
where $\pm$ denotes each detector,
$t$ is the time (in \unit{\micro\second}) after implantation,
$N_{\pm, 0}$ and $b_{\pm}$ denote the rate of ``good'' and ``background''
decay events,
$P_{\mu}(t) \in [-1, +1]$ is the muon spin-polarization,
and
$A_{0}$ is a proportionality constant
(\num{\sim 0.2} here).
The most essential part of \Cref{eq:counts} is $A_{0}P_{\mu}(t)$
and
it may be obtained directly by taking the \emph{asymmetry} of two counters:
\begin{equation}
  A_{0} P_{\mu}(t) = \frac{ \left [ N_{+}(t) - b_{+} \right ] - \alpha [ N_{-}(t) - b_{-} ] }{ \left [ N_{+}(t) - b_{+} \right ] + \alpha [ N_{-}(t) - b_{-} ] } ,
  \label{eq:asymmetry}
\end{equation}
where $\alpha \equiv N_{+, 0} / N_{-, 0}$ accounts for imperfections in the detector pair (e.g., different efficiencies, effective solid angles, etc.).
Important for this work is the temporal evolution of $P_{\mu}(t)$,
which contains all information about the local magnetic environment
below the sample's surface
(i.e., at the $\mu^{+}$ stopping site).
Fortunately,
$P_{\mu}(t)$ differs in each of Nb's superconducting states,
allowing us to quantify their volume fraction for
our set of measurement conditions.

For the simplest case of Nb's normal state,
where magnetic flux penetrates the sample's surface freely,
the \gls{musr} signal follows that of a
typical transverse-field measurement~\cite{2011-Yaouanc-MSR}:
\begin{equation}
  \label{eq:tf-gaussian}
  P_{\mathrm{GTF}} (t) = \exp \left [ -\frac{ \left ( \sigma t \right )^{2} }{2} \right ] \cos \left( \gamma_{\mu} \mu_{0} H_{0} t + \phi \right),
\end{equation}
where $t$ is the time after implantation,
$\sigma$ is the damping rate (from a Gaussian field distribution),
$\mu_{0} H_{0}$ is magnetic field at the $\mu^{+}$ stopping site
(dominated by the external applied field),
and
$\phi$ is a phase factor.
In non-superconducting Nb,
the term $\sigma \approx \qty{0.5}{\per\micro\second}$~\cite{2013-Grassellino-PRSTAB-16-062002,2017-Junginger-SST-30-125012,2018-Junginger-PRAB-21-032002},
causing minimal damping.
Conversely,
in Nb's vortex state,
where fluxoids form a periodic arrangement
with a broad distribution~\cite{1995-Brandt-RPP-58-1465,2003-Brandt-PRB-68-054506},
$\sigma$ is much larger
and
the signal is damped quickly
(i.e., within the first \qty{\sim 0.3}{\micro\second} following implantation).
In the opposite limit of Nb's Meissner state,
where all magnetic flux is expelled from the sample's interior,
the signal follows that of a so-called dynamic zero-field Kubo-Toyabe
function~\cite{1979-Hayano-PRB-20-850}:
\begin{multline}
  \label{eq:dynamic-kt}
  P_{\mathrm{DGKT}} (t) = P_{\mathrm{SGKT}} (t) \exp \left [ - \nu t \right ] + \\ \nu \int_{0}^{t} P_{\mathrm{DGKT}} (t - t^{\prime} ) P_{\mathrm{SGKT}}(t^{\prime}) \exp \left[ - \nu t^{\prime} \right]  \, \mathrm{d}t^{\prime} ,
\end{multline}
which is obtained from a static Kubo-Toyabe function
$P_{\mathrm{SGKT}}$~\cite{1967-Kubo-MRR-810}:
\begin{equation}
  \label{eq:static-kt}
  P_{\mathrm{SGKT}}(t) =
  \frac{1}{3} + \frac{2}{3} \left[1 - (\sigma t)^2 \right] \exp \left [- \frac{ \left ( \sigma t \right )^{2} }{2} \right ] ,
\end{equation}
where the local field is fluctuating
(e.g., from stochastic site-to-site ``hopping'' of $\mu^{+}$~\cite{2014-Karlsson-EPJH-39-303})
at a rate $\nu$,
typically \qty{\sim 0.7}{\per\micro\second} for \gls{srf} Nb~\cite{2013-Grassellino-PRSTAB-16-062002,2017-Junginger-SST-30-125013}.
Note that both \Cref{eq:dynamic-kt,eq:static-kt} assume the local field
distribution at the $\mu^+$ stopping site to be Gaussian,
in accord with related
studies~\cite{2013-Grassellino-PRSTAB-16-062002,2017-Junginger-SST-30-125012,2017-Junginger-SST-30-125013,2018-Junginger-PRAB-21-032002}.

For the present experiments,
the \gls{musr} signal is, in general,
described by a superposition of \Cref{eq:tf-gaussian,eq:dynamic-kt},
which may be written as:
\begin{equation}
  \label{eq:polarization}
  P_{\mu}(t) = f_{\mathrm{ZF}} P_{\mathrm{DGKT}}(t) + (1-f_{\mathrm{ZF}}) \sum_{i} f_{\mathrm{TF}, i}  P_{\mathrm{GTF}, i}(t) ,
\end{equation}
where $f_{\mathrm{ZF}} \in [0, 1]$ denotes the volume fraction of the
zero-field component,
which for superconducting Nb in an applied field
corresponds to its Meissner state,
and
$f_{\mathrm{TF}, i} \in [0, 1]$ represents the individual
transverse-field components
(e.g., normal state, vortex state, etc.),
subject to the constraint that $\sum_{i} f_{\mathrm{TF}, i} \equiv 1$.
Examples of this type of composite signal are shown in \Cref{fig:musr-data}.

In order to identify the $\mu_{0}H_\mathrm{vp}$ in each sample,
the evolution of $f_{\mathrm{ZF}}$ in
(monotonically increasing)
magnetic fields $\mu_{0}H_{0}$ up to \qty{\sim 260}{\milli\tesla}
was measured at the cryostat's base temperature
($T \approx \qty{2.7}{\kelvin}$)
following zero-field cooling.
Any depth-dependence was inferred from repeat measurements
using different moderator foil thicknesses (and $\mu^{+}$ beam energies).
Note that,
to ensure the accuracy of the applied fields reported for all superconducting states
(i.e., due to its geometric enhancement from flux-expulsion),
all values were derived from field calibration measurements conducted above $T_\mathrm{c}$
(i.e., at $T \geq \qty{10.5}{\kelvin}$ for the \gls{ltb} Nb
and
$T \geq \qty{20}{\kelvin}$ for the \gls{ss} bilayer),
where the \gls{musr} signal simply follows \Cref{eq:tf-gaussian}.
Specifically,
they were corrected using~\cite{2000-Brandt-Physica-332-99}:
\begin{equation}
  \mu_{0}H_{0} = \mu_{0}H_{0}^{\mathrm{NS}}\times\frac{1}{1-N} ,
  \label{eq:demagn-factor}
\end{equation}
where $\mu_{0}H_{0}^{\mathrm{NS}}$ is the (measured) applied field in the normal state
and
$N \approx 0.13$~\cite{2017-Junginger-SST-30-125012,2018-Junginger-PRAB-21-032002}
is the demagnetization factor for our samples
(see \Cref{sec:experiment:samples}).

\subsection{
  Samples
  \label{sec:experiment:samples}
}

The samples used in this study are identical to those employed in previous \gls{musr}
measurements on the first-flux-penetration field~\cite{2017-Junginger-SST-30-125012,2018-Junginger-PRAB-21-032002}.
For completeness,
we briefly restate their preparation details below.

Each Nb sample was cut from fine-grain Nb (Wah Chang Corporation) stock sheets with a \gls{rrr} \num{>150} 
and
machined into prolate ellipsoids
with a semi-major radius \qty{22.9}{\milli\meter}
and
semi-minor radii
\qty{9.0}{\milli\meter} for the other axes.
After machining,
the samples underwent \gls{bcp} to remove the surface's topmost
\qty{100}{\micro\meter} of the material
(see, e.g.,~\cite{2011-Ciovati-JAE-41-721}).
For one of the samples,
a typical \gls{ltb} ``recipe'' was followed where,
after degassing in vacuum at \qty{800}{\celsius} for \qty{4}{\hour},
the Nb ellipsoid was baked in vacuum at
\qty{120}{\celsius} for \qty{48}{\hour}~\cite{2004-Ciovati-JAP-96-1591}. A complementary magnetostatic characterization of the
\gls{ltb} sample can also be found elsewhere~\cite{2022-Turner-SR-12-5522}.
Microstructure analysis was conducted on identically prepared flat ``witness'' samples using \gls{sem}/\gls{edx} and \gls{sims}. The \gls{sem}/\gls{edx} analysis revealed no unexpected features or contaminations beyond typical observations for high-\gls{rrr} Nb exposed to air~\cite{2012-Antoine-Book}. 
\gls{sims} confirmed the expected increase in near-surface oxygen concentration following \gls{ltb} for depths \qty{< 20}{\nano\meter}, in accord with models for oxygen diffusion~\cite{2006-Ciovati-APL-89-022507,2024-Lechner-JAP-135-133902,2021-Lechner-APL-119-082601}.
For another sample,
a \qty{2}{\micro\meter} Nb$_3$Sn surface coating was applied
using a vapour diffusion~\cite{1962-Saur-N-49-127,1977-Arnolds-IEEETM-13-500}
procedure developed at
Cornell University~\cite{2014-Posen-PRSTAB-17-112001,2015-Posen-APL-106-082601}.
\Gls{sem} and \gls{afm} measurements on identically prepared ``witness'' samples revealed that the as-grown Nb$_3$Sn exhibits surface roughness comparable to their grain size, typically in the \unit{\micro\meter} range~\cite{2017-Posen-SST-30-033004}. Further tests on substrates 
prepared using different polishing techniques had no effect on the bilayer's surface roughness~\cite{2015-Eremeev-SRF-2015}.

\section{
  Results
  \label{sec:results}
 }

\begin{figure}
  \centering
  \includegraphics*[width=\columnwidth]{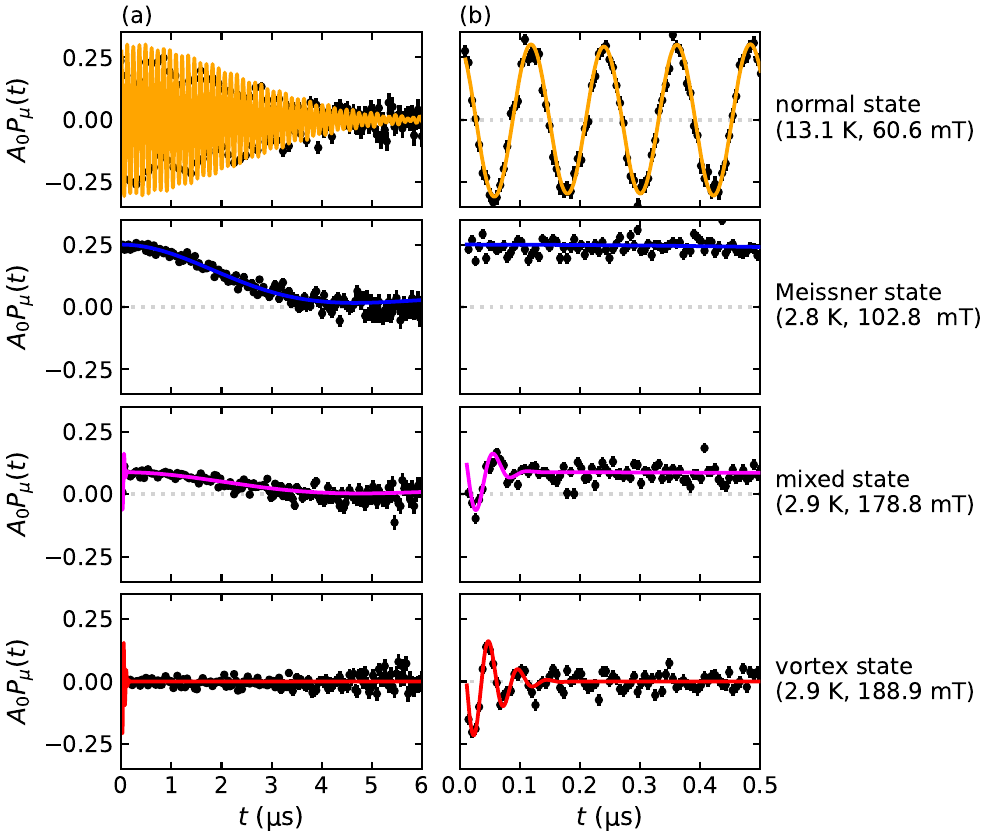}
  \caption{
  \label{fig:musr-data}
  Typical \gls{musr} data in surface-treated Nb,
  illustrating the evolution with applied magnetic field
  $\mu_{0}H_{\mathrm{0}}$.  Both panels \textbf{(a)} and \textbf{(b)} display the same dataset; however, \textbf{(b)} specifically represents the initial \qty{0.5}{\micro\second}.  The temperature $T$ and $\mu_{0}H_{0}$ for each superconducting state are detailed in the panel's inset.  The solid lines are fits to the data points using~\Cref{eq:polarization}.
  In the normal state,
  coherent spin-precession is observed with minimal damping,
  consistent with a narrow local field distribution.
  In the Meissner state,
  the applied field is completely screened
  and
  the local field is dominated by the host's \ch{^{93}Nb}
  nuclear spin,
  resulting in the characteristic (dynamic) ``zero-field''
  signal.
  At fields just above the vortex penetration field,
  a mixed signal with both zero- and transverse-field
  components is observed,
  the latter being strongly damped.
  Finally,
  in the vortex state,
  the broad field distribution causes strong damping of
  the transverse-field response.
  }
\end{figure}

\Cref{fig:musr-data} depicts typical time-differential \gls{musr}
data in one of our samples
(\gls{ltb} Nb~\cite{2004-Ciovati-JAP-96-1591}),
showing the contrast in signals for different material states.
In the normal state,
where the local field at the $\mu^{+}$ stopping sites is dominated
by the (transverse) applied magnetic field $\mu_{0} H_{0}$,
coherent spin-precession is observed with minimal damping,
consistent with a narrow field distribution.
By contrast,
in the Meissner state,
$\mu_{0}H_{0}$ is completely screened~\footnote{The range of implanted $\mu^{+}$ greatly exceeds Nb's London penetration depth (\qty{\sim 29}{\nano\meter}~\cite{2023-McFadden-PRA-19-044018}), leading to the absence of magnetic flux density for depths greater than \qty{\sim 100}{\nm}.}
and
the local field is dominated, in part,
by Nb's \qty{100}{\percent} abundant \ch{^{93}Nb} nuclear spins,
resulting in the characteristic (dynamic) ``zero-field''
signal~\cite{1979-Hayano-PRB-20-850}.
At applied fields just above the vortex penetration field $\mu_{0} H_{\mathrm{vp}}$,
a mixed signal with both zero- and transverse-field
components is observed,
the latter being strongly damped.
As the field is further increased beyond $\mu_{0} H_{\mathrm{vp}}$,
the sample fully enters a vortex state,
where the signal resembles that of the normal state,
but the broad field distribution from the vortex lattice
causes strong damping of the spin-polarization.
Similar behavior is observed at other implantation ranges,
as well as in the Nb$_3$Sn(\qty{2}{\micro\meter})/Nb sample
(not shown).
In line with the experiment's description in \Cref{sec:experiment},
these observations can be quantified through fits to \Cref{eq:polarization},
yielding good agreement with the data in all cases
(typical reduced-$\chi^{2} \approx 1.07$).
A subset of the fit results are shown in \Cref{fig:musr-data}.

\begin{figure}
  \centering
  \includegraphics*[width=\columnwidth]{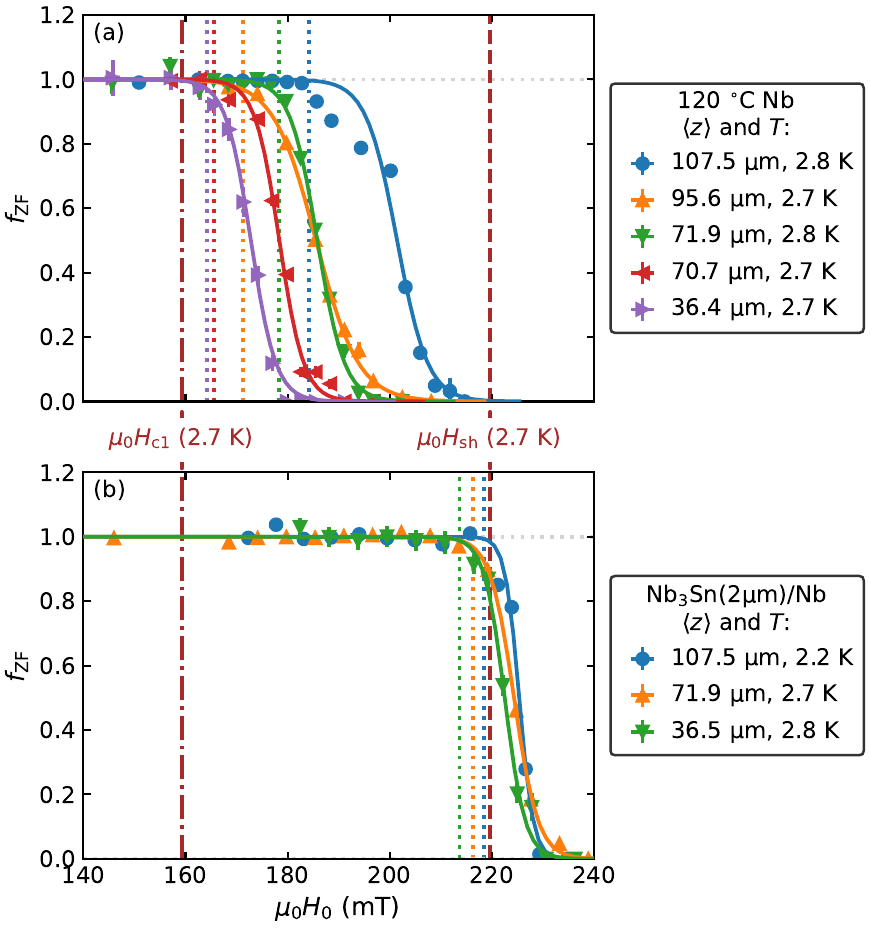}
  \caption{
  \label{fig:volume-fractions}
  Volume fraction of the zero-field \gls{musr} signal $f_{\mathrm{ZF}}$
  at different applied magnetic field $\mu_{0} H_{0}$
  and
  mean $\mu^{+}$ implantation depths $\langle z \rangle$. The $\langle z \rangle$ and measurements temperature $T$ are mentioned in the figure's inset.
  The solid colored lines denotes fits to a logistic function,
  intended to guide the eye. The $\mu_{0}H_{\mathrm{vp}}$ at each $\langle z \rangle$ are shown using colored dotted vertical lines.
  Vertical dotted-dashed and dashed brown lines are included to mark Nb's
  lower critical field $\mu_{0} H_\mathrm{c1}$
  and
  superheating field $\mu_{0} H_{\mathrm{sh}}$ at \qty{2.7}{\kelvin}.
  Note that \ch{Nb_3Sn} has a considerably smaller $\mu_{0} H_\mathrm{c1} = \qty{25.0 \pm 1.4}{\milli\tesla}$~\cite{2015-Posen-PRL-115-047001} compared to the fields shown here.
  \textbf{(a)}:
  In \gls{ltb} Nb, the vortex penetration field $\mu_{0} H_{\mathrm{vp}}$
  is comparable to $\mu_{0} H_{\mathrm{c1}}$,
  but shows a strong $\langle z \rangle$-dependence,
  increasing with increasing $\langle z \rangle$.
  \textbf{(b)}:
  In Nb$_3$Sn(\qty{2}{\micro\meter})/Nb,
  $\mu_{0}H_{\mathrm{vp}}$ is $\langle z \rangle$-independent
  and close to $\mu_{0} H_{\mathrm{sh}}$.
  }
\end{figure}

To aid in identifying $\mu_{0} H_{\mathrm{vp}}$ and its depth-dependence,
we plot the measured $f_{\mathrm{ZF}}$s vs.\ $\mu_{0}H_{0}$
for each Nb sample in \Cref{fig:volume-fractions}.
The resulting ``curves'' all have a sigmoid-like shape,
where $f_{\mathrm{ZF}} = 1$ for $\mu_{0}H_{0} < \mu_{0}H_{\mathrm{vp}}$,
with $f_{\mathrm{ZF}}$ decreasing rapidly towards zero
once $\mu_{0} H_{0} \geq \mu_{0}H_{\mathrm{vp}}$.
This behavior is phenomenologically captured by a logistic function:
\begin{equation}
  \label{eq:logistic}
  f_{\mathrm{ZF}} ( \mu_{0} H_{0} ) \approx \frac{ f_{0} }{ \exp \left [ -k \left ( \mu_{0} H_{0} - \mu_{0} H_\mathrm{m} \right ) \right ] + 1},
\end{equation}
where
$f_{0}$ is the curve's height,
$k$ denotes the ``steepness'' of the transition,
and
$\mu_{0} H_\mathrm{m}$ represents its midpoint.
Fits of the measured volume fractions to \Cref{eq:logistic}
are shown in \Cref{fig:volume-fractions} as guides to the eye.

It is clear from \Cref{fig:volume-fractions} that $\mu_{0} H_{\mathrm{vp}}$'s
depth-dependence is quite different for \gls{ltb} and coated Nb.
In the \gls{ltb} sample,
flux-penetration is detected just above Nb's
$\mu_{0}H_\mathrm{c1}$,
with the onset pushed to higher $\mu_{0}H_{0}$s for larger $\mu^{+}$ ranges.
Similarly,
the field span of this ``transition''
(i.e., going from zero- to full-flux-penetration)
also increases slightly deeper below the surface.
By contrast,
in the Nb$_3$Sn(\qty{2}{\micro\meter})/Nb,
first-flux-penetration onsets close to
Nb's $\mu_{0} H_{\mathrm{sh}}$,
with this value
and the transition's width both virtually
unaffected by $\mu^{+}$'s proximity to the sample's surface.
Qualitatively,
this disparity between the Nb treatments indicates that different mechanisms
are likely responsible for determining each sample's $\mu_{0}H_{\mathrm{vp}}$.

\begin{figure}
  \centering
  \includegraphics*[width=\columnwidth]{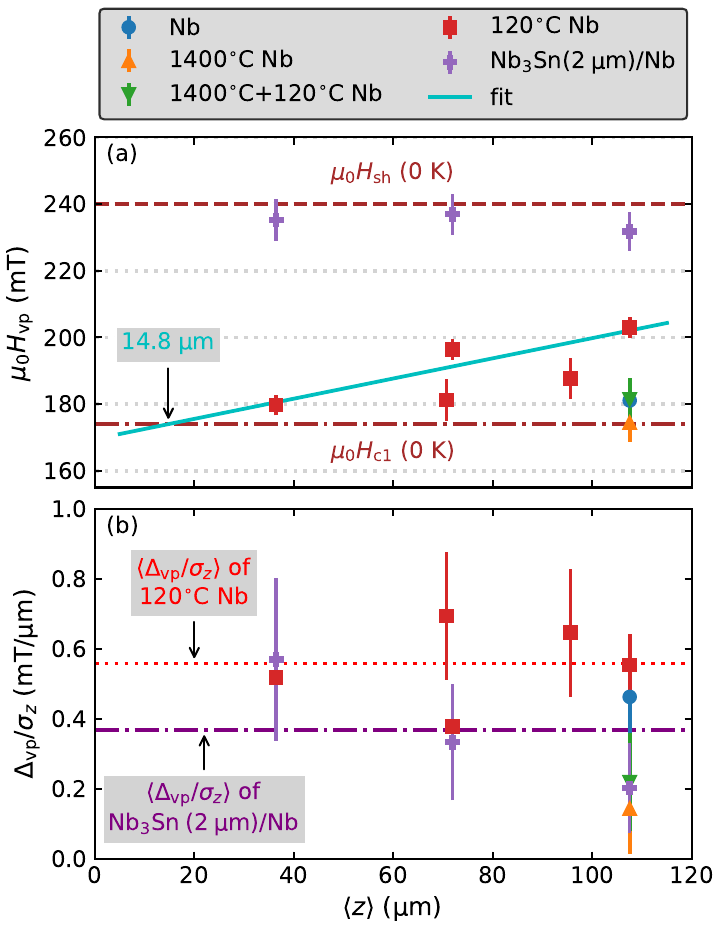}
  \caption{
  \label{fig:vortex-penetration-fields}
  Summary of the first-flux penetration measurements
  at different mean $\mu^{+}$ implantation depths $\langle z \rangle$
  for \gls{ltb} Nb and Nb$_3$Sn(\qty{2}{\micro\meter})/Nb.
  For comparison,
  we include re-analyzed results for additional \gls{srf} Nb treatments
  (originally reported elsewhere~\cite{2017-Junginger-SST-30-125012,2018-Junginger-PRAB-21-032002}).
  \textbf{(a)}:
  Measured vortex penetration fields $\mu_{0} H_{\mathrm{vp}}$.
  The horizontal dashed and dotted-dashed brown lines denote Nb's
  superheating field $\mu_{0} H_{\mathrm{sh}}$
  and
  lower critical field $\mu_{0} H_\mathrm{c1}$,
  respectively.
  In Nb$_3$Sn(\qty{2}{\micro\meter})/Nb,
  $\mu_{0}H_{\mathrm{vp}}$ is $\langle z \rangle$-independent
  and close to $\mu_{0} H_{\mathrm{sh}}$,
  whereas $\mu_{0} H_{\mathrm{vp}} \approx \mu_{0} H_\mathrm{c1}$ in \gls{ltb} Nb,
  increasing modestly with increasing $\langle z \rangle$.
  The other surface-treatments have $\mu_{0} H_{\mathrm{vp}}$s that are
  similarly close to $\mu_{0} H_\mathrm{c1}$.  The cyan color solid line represents the ``straight line'' fit applied to the \gls{ltb} data, providing an estimate of the depth where $\mu_{0} H_{\mathrm{vp}} = \mu_{0} H_\mathrm{c1}$.
  \textbf{(b)}:
  Measured Meissner-vortex transition ``widths'' $\Delta \mu_{0} H_{\mathrm{vp}}$, divided by $\sigma_{z}$, confirm that the stopping distributions are not significantly different.
  The dotted and dash-dotted horizontal lines are the average values of $\Delta_{\mathrm{vp}}/\sigma_{z}$, $\langle \Delta_{\mathrm{vp}}/\sigma_{z} \rangle$ for \gls{ltb} and Nb$_3$Sn(\qty{2}{\micro\meter})/Nb samples, respectively.
  }
\end{figure}

To quantify these differences explicitly,
we use a non-parametric approach to identify $\mu_{0}H_{\mathrm{vp}}$
for each ``curve'' in \Cref{fig:volume-fractions}.
Noting that each $f_{\mathrm{ZF}}$ has a statistical uncertainty of \qty{\sim 4}{\percent} (determined from fitting),
we define $\mu_{0} H_{\mathrm{vp}}$ to correspond to the applied field where
where $f_{\mathrm{ZF}} \leq 0.96$
(i.e., the very onset of flux-penetration).
Due to the finite ``sampling'' of our measurements,
this field is estimated as midpoint between the pair of $f_{\mathrm{ZF}}$s
on either side of the threshold
criteria~\footnote{While simple, this approach introduces an additional systematic uncertainty to the assignment of $\mu_{0}H_{\mathrm{vp}}$. We estimate this quantity as one-sixth the distance between the field points, such that $\pm 3$ standard deviations covers the full span of the ``uncertain'' region with \qty{\sim 100}{\percent} probability.}.
These values are marked graphically in \Cref{fig:volume-fractions} by vertical dotted colored lines.
Similarly,
we take the width of the zero- to full-flux-penetration ``transition''
$\Delta_{\mathrm{vp}}$
to be the field range where
$\qty{4}{\percent} \leq f_{\mathrm{ZF}} \leq \qty{96}{\percent}$.
To correct for any influence from the finite span of the $\mu^{+}$ stopping profile
we additionally normalize the $\Delta_{\mathrm{vp}}$s by the width $\sigma_{z}$
of the implantation distribution, which varies for different beam implantation
conditions (see~\Cref{fig:stopping-profiles}).
For each of our samples,
the dependence of both $\mu_{0}H_{\mathrm{vp}}$ and $\Delta_{\mathrm{vp}}/\sigma_{z}$
on the implanted $\mu^{+}$ range $\langle z \rangle$ is shown in
\Cref{fig:vortex-penetration-fields}.
For comparison,
measured values for additional surface-treated samples~\cite{2017-Junginger-SST-30-125012,2018-Junginger-PRAB-21-032002}
have also been included~\footnote{To ensure the self-consistency of our $\mu_{0}H_{\mathrm{vp}}$ comparison in \Cref{fig:vortex-penetration-fields}, following the approach described in \Cref{sec:experiment}, we re-analyzed the raw \gls{musr} data for select samples originally reported in Refs.~\citenum{2017-Junginger-SST-30-125012,2018-Junginger-PRAB-21-032002}. As anticipated, our updated values are in good agreement with those in the original reports.}.
Lastly,
to facilitate comparison between the measured $\mu_{0}H_{\mathrm{vp}}$s,
we correct for minor temperature differences
and
extrapolate our values to \qty{0}{\kelvin}
using the empirical relation~\cite{1996-Tinkham-IS-2}:
\begin{equation}
  \mu_{0}H_\mathrm{vp}(T) \approx \mu_{0}H_\mathrm{vp}(\qty{0}{\kelvin}) \left[ 1 - \left(\frac{T}{T_{\mathrm{c}}} \right)^2 \right],
  \label{eq:temp-correction}
\end{equation}
where $T$ is the absolute temperature,
$T_{\mathrm{c}} \approx \qty{9.25}{\kelvin}$ is Nb's critical temperature~\cite{1966-Finnemore-PR-149-231},
and
$\mu_{0}H_\mathrm{vp}(\qty{0}{\kelvin})$ is the value at absolute zero.

\section{
  Discussion
  \label{sec:discussion}
 }

Consistent with our main observations in \Cref{fig:volume-fractions}
for the vortex penetration field,
\Cref{fig:vortex-penetration-fields}(a) shows that $\mu_{0} H_\mathrm{vp}$s in Nb$_3$Sn(\qty{2}{\micro\meter})/Nb
(extrapolated to \qty{0}{\kelvin})
remain  depth-independent
with an average value of \qty{234.5 \pm 3.5}{\milli\tesla},
in excellent agreement with Nb's $\mu_{0}H_{\mathrm{sh}}(\qty{0}{\kelvin}) \approx \qty{240}{\milli\tesla}$~\cite{2015-Posen-PRL-115-047001,2017-Junginger-SST-30-125012}.
That the flux penetration occurs in such close proximity to the superheating field
is strong evidence for the presence of an interface barrier
at the \gls{ss} boundary, similar to a \gls{bl} surface energy barrier~\cite{1964-Bean-PRL-12-14},
as anticipated by the theoretical framework
for superconducting multilayers~\cite{2017-Kubo-SST-30-023001,2014-Kubo-APL-104-032603}.
Conversely,
in \gls{ltb} Nb the $\mu_{0}H_{\mathrm{vp}}$s are much lower,
remaining close to (but slightly above) $\mu_{0}H_{\mathrm{c1}}$
for all measurements.
In this case, however, a modest depth-dependence is observed,
with $\mu_{0}H_{\mathrm{vp}}$
increasing gradually with increasing $\langle z \rangle$.
That these details coincide is significant,
as it sets \gls{ltb} apart from other surface treatments,
where flux-penetration occurs at Nb's lower critical field.
We note that, though small,
the observation of such a depth-dependence is inconsistent
with a surface energy barrier being solely responsible for pushing
$\mu_{0}H_{\mathrm{vp}} > \mu_{0}H_{\mathrm{c1}}$~\footnote{A similar argument can be made for the Nb$_3$Sn(\qty{2}{\micro\meter})/Nb sample (i.e., that flux pinning --- particularly from the thin \ch{Nb_3Sn} layer ---
may contribute to the increased $\mu_{0}H_{\mathrm{vp}}$ we observe).
We point out, however, that the observed lack of depth-dependence in our data, along with the magnitude in which $\mu_{0}H_{\mathrm{vp}}$ exceeds Nb's $\mu_{0}H_{\mathrm{c1}}$, suggest that pinning is not the dominant mechanism for elevating the vortex penetration field in the \gls{ss} bilayer.}.
An alternative possibility is the presence of pinnning centers
localized near \gls{ltb} Nb's surface,
which has been suggested previously~\cite{2018-Junginger-PRAB-21-032002,2022-Turner-SR-12-5522}.
In such a case, $\mu_{0} H_\mathrm{vp}$ closely approximates $\mu_{0} H_\mathrm{c1}$
due to the presence of pinning,
which diminishes towards the sample's center,
resulting in delayed flux penetration
(i.e., the pinning centers act as supplemental flux ``blockades,''
providing resistance to the free motion of the fluxoids, which would
otherwise uniformly distribute throughout the sample)
~\cite{2018-Junginger-PRAB-21-032002}.
Independent of the mechanistic details,
from our data we identify the length scale over which flux-penetration is retarded.
Using a simple linear fit to the measured values,
we find that $\mu_{0}H_{\mathrm{vp}} \approx \mu_{0}H_{\mathrm{c1}}$
at a mean depth of \qty{\sim 14}{\micro\meter},
characterizing the distance in which it is delayed for our sample geometry. Note that, there is a significant proportion of muons that stop significantly closer to the surface at this average depth.
We shall return the implications of this quantity later on.

Further insight into the flux-penetration mechanism for the \gls{ltb}
and \gls{ss} samples can be gleaned from
the (normalized) flux-entry ``transition''
widths $\Delta_{\mathrm{vp}}/\sigma_{z}$,
which are shown in \Cref{fig:vortex-penetration-fields}(b).
The span from first- to full-flux penetration provides a measure for the ``haste''
in which vortices nucleate through the probe $\mu^{+}$ stopping depths where,
in the presence of near-surface pinning,
we expect that the Meissner-vortex transition becomes ``extended'' to a larger range
of applied fields
(i.e., the presence of pinning centers delays full-flux penetration).
Thus,
we suggest that $\Delta_{\mathrm{vp}}/\sigma_{z}$ serves as a proxy for each treatment's pinning strength.
Indeed, we observe that the values for \gls{ltb} Nb are all similar, which also suggests that the pinning strength appears to be depth-independent and exceeds the values of all other surface preparations shown in~\Cref{fig:vortex-penetration-fields}. However, given the relatively large uncertainty in each measurement, drawing firm conclusions about treatment-specific differences is challenging.
This is clear from their average values,
which turned out to be
\qty{0.56 \pm 0.15}{\milli\tesla/\micro\meter} (for \gls{ltb})
and
\qty{0.37 \pm 0.18}{\milli\tesla/\micro\meter} (for Nb$_3$Sn(\qty{2}{\micro\meter})/Nb),
respectively.
Interestingly,
both quantities are comparable to that of Nb in the absence of any treatment
(\qty{0.46 \pm 0.14}{\milli\tesla/\micro\meter}),
which is larger than both the values obtained for \qty{1400}{\celsius} annealing
(\qty{0.14 \pm 0.13}{\milli\tesla/\micro\meter}),
as well as
\qty{1400}{\celsius} annealing + \qty{120}{\celsius} \gls{ltb}
(\qty{0.22 \pm 0.14}{\milli\tesla/\micro\meter}).
We note that,
in line with our expectations for $\Delta_{\mathrm{vp}} / \sigma_{z}$,
its value is minimized for the \qty{1400}{\celsius} treatment,
which is known to release virtually all pinning~\cite{2018-Junginger-PRAB-21-032002}.
Thus,
although not as conclusive as the $\mu_{0}H_{\mathrm{vp}}$ measurements,
the large values for \gls{ltb} insinuate that pinning is strongest for
this treatment.
With these results in mind,
we will explore their implications in the remaining discussion.

First,
we consider the \gls{ss} bilayer Nb$_3$Sn(\qty{2}{\micro\meter})/Nb,
whose high $\mu_{0}H_{\mathrm{vp}}$ is favorable for technical applications
requiring operation in a flux-free state
(e.g., \gls{srf} cavities).
In fact,
\gls{dc} measurements of first-flux-penetration using a Hall probe magnetometer on a
\qty{1.3}{\giga\hertz} single-cell \gls{srf} cavity of similar composition
are in good agreement with our result~\cite{2015-Posen-PRL-115-047001},
with similar $\mu_{0}H_{\mathrm{vp}}$ values reported for surface coatings other than Nb$_3$Sn,
such as \gls{hpcvd} MgB$_2$~\cite{2017-Junginger-SST-30-125012}.
The combination of our work and related studies~\cite{2017-Junginger-SST-30-125012,2018-Junginger-PRAB-21-032002,2015-Posen-PRL-115-047001}
provides compelling evidence for the use of
\gls{ss} bilayers as an empirical means for increasing $\mu_{0}H_{\mathrm{vp}}$,
consistent with earlier observations using \gls{musr}~\cite{2017-Junginger-SST-30-125012,2018-Junginger-PRAB-21-032002}.
For further insight into why such a treatment is so effective at enhancing $\mu_{0}H_{\mathrm{vp}}$,
we must consider the theory of superconducting multilayers
(see, e.g.,~\cite{2017-Kubo-SST-30-023001}),
which we shall briefly outline below.

In bilayer superconductors,
by analogy with the \gls{bl} barrier at the surface of
``bulk'' superconductors~\cite{1964-Bean-PRL-12-14},
the discontinuity in each material's (coupled) electromagnetic response at the \gls{ss} interface
is responsible for creating a second (sub-surface) barrier
that impedes flux penetration~\cite{2014-Kubo-APL-104-032603}.
Specifically,
electromagnetic continuity across the \gls{ss} boundary creates a ``coupling'' between the
layer's properties,
leading to marked deviations from the lone material's native behavior
when the surface layer penetration depth is larger than the substrate.
Microscopically,
this is predicted to manifest in the heterostructure's Meissner screening profile
with a distinct bipartite form~\cite{2014-Kubo-APL-104-032603},
which was recently confirmed experimentally for
$\mathrm{Nb_{1-x}Ti_{x}N}$(\qtyrange{50}{160}{\nano\meter})/Nb
samples~\cite{2024-Asaduzzaman-SST-37-025002}.
A weaker Meissner screening current is observed, which,
as a corollary,
provides enhanced protection against flux nucleation
(see, e.g.,~\cite{2017-Kubo-SST-30-023001}).
While the $\mu_{0}H_{\mathrm{vp}} \approx \mu_{0}H_{\mathrm{sh}}$ we observe
is in good agreement with this prediction,
the theory also suggests that $\mu_{0}H_{\mathrm{vp}}$ can be improved
further still through optimizing the Nb$_3$Sn coating's thickness.
This enhancement is achievable by ensuring a flux-free surface layer and enabling superheating in both the surface and substrate layers. Achieving this involves precise adjustments to the thickness of the surface layer and the presence of an interface barrier at the \gls{ss} boundary for the substrate layer.
For an \gls{ss} structure,
the optimum thickness of the surface layer $d_\mathrm{s}^\mathrm{opt}$
is defined by~\cite{2017-Kubo-SST-30-023001}:
\begin{multline}
  d_\mathrm{s}^\mathrm{opt} = \lambda_\mathrm{s} \log \left [\dfrac{\lambda_\mathrm{s} H_\mathrm{sh}^\mathrm{s}}{\left(\lambda_\mathrm{s} + \lambda_\mathrm{sub}\right) H_\mathrm{sh}^\mathrm{sub}} \right. + \\
    \left. \sqrt{ \left(\dfrac{\lambda_\mathrm{s} H_\mathrm{sh}^\mathrm{s}}{(\lambda_\mathrm{s} + \lambda_\mathrm{sub})H_\mathrm{sh}^\mathrm{sub}}\right)^2 + \left( \dfrac{\lambda_\mathrm{s} - \lambda_\mathrm{sub}}{\lambda_\mathrm{s} + \lambda_\mathrm{sub}} \right) }  \right] ,
  \label{eq:optimum-thickness-surface}
\end{multline}
where $\lambda_i$ and $H_\mathrm{sh}^{i}$ denote the magnetic penetration depth
and
superheating field
for the surface ($i$ = s) and substrate ($i$ = sub) layers.
Using literature values for these quantities
($\lambda_\mathrm{Nb_3Sn} = \qty{173 \pm 32}{\nm}$~\cite{2019-Keckert-SST-32-075004,2015-Posen-PRL-115-047001};
$B_\mathrm{sh}^\mathrm{Nb_3Sn} = \qty{430 \pm 110}{\milli\tesla}$~\cite{2015-Posen-PRL-115-047001,2019-Keckert-SST-32-075004};
$\lambda_\mathrm{Nb} = \qty{29.01 \pm 0.10}{\nm}$~\cite{2023-McFadden-PRA-19-044018};
and
$B_\mathrm{sh}^\mathrm{Nb} = \qty{237 \pm 29}{\milli\tesla}$~\cite{2015-Posen-PRL-115-047001,2017-Junginger-SST-30-125012}),
\Cref{eq:optimum-thickness-surface} yields
$d_\mathrm{s}^\mathrm{opt} = \qty{210 \pm 60}{\nm}$
or,
equivalently,
$\num{\sim 1.2} \lambda_\mathrm{Nb_3Sn}$.
Similar predictions have been made for
$\mathrm{Nb_{1-x}Ti_{x}N}$/Nb~\cite{2024-Asaduzzaman-SST-37-025002}.
It would be interesting to test these explicitly,
for example, using the experimental formalism employed in this work.
Investigating this phenomena in closely related \gls{sis} heterostructures
would also be fruitful,
as they offer similar means of enhancing $\mu_{0}H_{\mathrm{vp}}$~\cite{2017-Kubo-SST-30-023001}.

As a close to our discussion of the \ch{Nb_3Sn}/Nb bilayer,
it is interesting to consider its synthesis.
We note that in our sample,
as is common for heterostructures prepared by thermal diffusion,
the composition of the \ch{Nb_3Sn}/Nb interface deviates appreciably
from each layer's respective ``bulk''~\cite{2017-Posen-SST-30-033004,2018-Trenikhina-SST-31-015004}.
Specifically,
within the first few hundred nanometers from the heterojunction,
a localized \ch{Sn} deficiency (enhancement) is present in the Nb$_3$Sn (Nb) layers,
with the former known to lower Nb$_3$Sn's $T_{\mathrm{c}}$,
making the region a poor superconductor~\cite{2017-Posen-SST-30-033004,2018-Trenikhina-SST-31-015004}.
The presence of such inhomogeneities,
however,
do not meaningfully impact $\mu_{0}H_{\mathrm{vp}}$,
as indicated by it's agreement with Nb's $\mu_{0}H_{\mathrm{sh}}$.
Testing the extent in which this remains true,
for example,
on samples with extended defect regions,
would be interesting.
Similarly,
these nanoscale inhomogeneities at the \gls{ss} interface could be examined
directly using a depth-resolved technique such as
\gls{le-musr}~\cite{2004-Bakule-CP-45-203,2004-Morenzoni-JPCM-16-S4583},
with the caveat that the Nb$_3$Sn layer thickness must be compatible with
the technique's spatial sensitivity
(i.e., subsurface depths typically \qty{< 150}{\nano\meter}).
Finally,
given the presence of a secondary energy barrier at the \gls{ss} interface,
we suggest that incorporating \gls{ss} bilayers into \gls{srf} cavity structures
holds great promise for surpassing the inherent limitations of current Nb cavity technology
(i.e., enabling higher accelerating gradients and enhanced performance in particle accelerators).

We now turn our attention to the
\qty{120}{\celsius} \gls{ltb} treatment~\cite{2004-Ciovati-JAP-96-1591},
which is well-known in \gls{srf} applications for its ability to alleviate the so-called
\gls{hfqs} ``problem''~\cite{2009-Padamsee-RFSSTA,2017-Padamsee-SST-30-053003},
where a rapid decrease in a cavity's \gls{qf} occurs as the
peak surface magnetic field exceeds \qty{\sim 100}{\milli\tesla}~\footnote{Note that, in regards to \gls{hfqs} mitigation, the most effective \gls{ltb} treatments generally include an \gls{ep} step in place of the \gls{bcp} used in this work.}.
True to this fashion,
our finding of $\mu_{0}H_{\mathrm{vp}}$s in excess of Nb's $\mu_{0}H_{\mathrm{c1}}$
underscores its utility in this domain;
however,
the treatment's depth-dependent $\mu_{0}H_{\mathrm{vp}}$
and
relatively large $\Delta_{\mathrm{vp}} / \sigma_{z}$
make it unique among the comparison treatments reported here.
As mentioned above,
near-surface pinning of the flux-lines provides the most likely explanation for these facts.
The observed delay in flux-penetration would then arise from the pinning centers
acting as ``supplementary barriers,''
impeding the movement of vortices from the edges of the sample to the center~\cite{2018-Junginger-PRAB-21-032002}.
The relatively large Meissner-vortex transition ``widths'' observed here
also support this interpretation.
It has been suggested by others that  material inhomogeneities,
such as interstitial oxygen or hydrogen precipitates,
may dominate the pinning mechanism~\cite{2012-Dhavale-SST-25-065014}.
For further insight into the matter,
it is instructive to consider some of the treatment's finer details,
which we do below.

During \gls{ltb},
the heat treatment induces changes to Nb's superfluid density
in its outermost nanoscale region through the dissolution and diffusion of oxygen
originating from the metal's native surface oxide~\cite{2004-Ciovati-JAP-96-1591}.
This alteration is believed to result in a ``dirty'' region localized near Nb's surface
(i.e., the first \qty{\sim 50}{\nano\meter}),
as explained by an oxygen diffusion model~\cite{2006-Ciovati-APL-89-022507}.
This length scale aligns well other work,
including an experiment that used repeat \ch{HF} ``rinses'' to remove the topmost
\qty{\sim 50}{\nano\meter}
and (essentially) restore the \gls{hfqs} following \gls{ltb}~\cite{2013-Romanenko-PRSTAB-16-012001}.
Similarly,
an increase of the ratio of the upper and surface critical fields after baking
was explained by the presence of an impurity layer of thickness smaller than
Nb's
coherence length~\cite{2005-Casalbuoni-NIMA-538-45},
which is of similar magnitude.
Other work on related treatments have also found similar
results~\cite{2019-Romanenko-SRF-866,2021-Lechner-APL-119-082601},
and the first \qty{\sim 10}{\nano\meter} may be particularly enriched with
interstitial oxygen~\cite{2008-Delheusy-APL-92-101911}.
It has been suggested that the \gls{ltb} effect
(i.e., \gls{hfqs} mitigation)
is due to the strong suppression of
hydride precipitation~\cite{2013-Romanenko-SST-26-035003},
as oxygen efficiently traps interstitial (or ``free'') hydrogen
that has accumulated during standard chemical treatments,
such as \gls{bcp} or \gls{ep}~\cite{2013-Romanenko-SST-26-035003,2013-Denise-SST-26-105003,2019-Romanenko-SRF-866}.
Indeed, \gls{ltb} has been linked to changes in the vacancy structure in Nb's
near-surface region~\cite{2013-Romanenko-APL-102-232601,2015-Trenikhina-JAP-117-154507},
supporting the prevailing idea that nanoscale niobium hydrides
cause the \gls{hfqs}~\cite{2013-Romanenko-SST-26-035003}.
These works all point to the importance of surface defects,
especially those closest to the surface.

Within the \qty{\sim 50}{\nano\meter} ``dirty'' region,
it is expected that quantities sensitive to the density of (nonmagnetic) scattering centers
(e.g., the carrier mean-free-path, the magnetic penetration depth, etc.)
be altered from their (clean-limit) ``bulk'' values.
As the doping is likely inhomogeneous over this length scale,
a similar character may be imparted on dependent quantities.
Early experimental results seemed to favor this possibility~\cite{2014-Romanenko-APL-104-072601},
with other authors suggesting a strong likeness of \gls{ltb} Nb
to an ``effective'' \gls{ss} bilayer~\cite{2017-Kubo-SST-30-023001}.
Subsequent experiments,
however,
have shown that such a distinction is far from clear,
with both a recent a commentary~\cite{2024-Ryan-APL-124-086101}
and
a separate \gls{le-musr} experiment~\cite{2023-McFadden-PRA-19-044018}
showing that the effects are homogeneous over subsurface depths spanning
\qtyrange{\sim 10}{\sim 160}{\nano\meter}.
Such a finding was rather surprising,
given the aforementioned related
work~\cite{2006-Ciovati-APL-89-022507,2013-Romanenko-PRSTAB-16-012001,2005-Casalbuoni-NIMA-538-45}
and that doping from the closely related nitrogen infusion
treatment~\cite{2017-Grassellino-SST-30-094004}
yields inhomogeneous superconducting properties
over the same length scale~\cite{2020-Checchin-APL-117-032601}.
We point out,
however,
that the observed electromagnetic response for \gls{ltb}~\cite{2024-Ryan-APL-124-086101,2023-McFadden-PRA-19-044018}
is consistent with the absence of an interface energy barrier preventing
flux-penetration~\cite{2019-Ngampruetikorn-PRR-1-012015,2020-Checchin-APL-117-032601},
in line with our present findings.

As alluded above,
the ``dirty'' nature of \gls{ltb} provides an ample environment for pinning centers,
which can serve as seeds for flux penetration.
While other experiments are clear on their surface proximity,
their \qty{\sim 50}{\nano\meter} localization is quite different from
the micrometer depth-dependence we observe for $\mu_{0}H_{\mathrm{vp}}$,
which warrants further consideration.
As is shown in \Cref{fig:vortex-penetration-fields}(a),
\gls{ltb}['s] $\mu_{0}H_{\mathrm{vp}}$ varies approximately linearly in $\langle z \rangle$.
From a fit to a function of the form
\begin{equation*}
  \mu_{0}H_{\mathrm{vp}}( \langle z \rangle ) \approx A + B \langle z \rangle ,
\end{equation*}
we parameterize this trend,
but postulate that Nb's $\mu_{0}H_{\mathrm{c1}}$ acts as the floor for
$\mu_{0}H_{\mathrm{vp}}$.
Upon equating the two relations,
we find that $\mu_{0}H_{\mathrm{vp}}( \langle z \rangle ) \approx \mu_{0} H_{\mathrm{c1}}$
when $\langle z \rangle \approx \qty{14}{\micro\meter}$.
Clearly,
this scale is considerably larger than the ``dirty'' region's
extent~\cite{2006-Ciovati-APL-89-022507,2013-Romanenko-PRSTAB-16-012001,2005-Casalbuoni-NIMA-538-45}.
We argue that,
despite the \gls{ltb} effect being confined to the very near surface,
this ``layer'' could introduce pinning over a \unit{\micro\meter} length scale in
an ellipsoidal geometry.
This is a consequence of the fact that,
even if flux lines penetrate further into the material,
they must both enter and exit through the ``dirty'' region.
The pinning strength is directly influenced by the flux line's
path length through this volume,
which in the case of an ellipsoid is minimized for the straight path along its equator,
but maximized for a (curved) trajectory close to the surface
(see \Cref{fig:experiment-sketch}).
Indeed,
magnetometry studies demonstrate that \gls{ltb} can significantly alter
the pinning characteristics in this geometry~\cite{2022-Turner-SR-12-5522}.
We emphasize that an ellipsoidal geometry is the ideal means for probing intrinsic pinning effects,
as opposed to other sample forms (e.g., rectangular prisms)
where additional geometric effects are present~\cite{2000-Brandt-Physica-332-99}.
In line with the generally accepted view that \gls{ltb} changes the
concentration of pinning centers
(i.e., from the redistribution of near-surface defects)~\cite{2010-Ciovati-PRSTAB-13-022002,2013-Romanenko-PRSTAB-16-012001,2020-Wenskat-SR-10-8300,2021-Lechner-APL-119-082601},
our identification of a length scale associated with flux pinning
may prove useful in further refining their microscopic distribution.
In the future,
it would be interesting to use this finding as a constraint for
simulations of flux-entry in ellipsoidal geometries
(see, e.g., Ref.~\citenum{1983-Brandt-JLTP-53-41}).

In terms of \gls{srf} cavity performance,
it is important to highlight that
our investigation on the
\gls{ltb} ``dirty'' layer cannot explain situations where cavities
in \gls{rf} operation exhibit $\mu_{0}H_{\mathrm{vp}}$s above $\mu_{0}H_{\mathrm{c1}}$, reaching values as high as
\qty{\sim 190}{\milli\tesla}~\cite{2011-Dhavale-AIPCP-1352-119,2012-Dhavale-SST-25-065014},
equivalent to accelerating gradients of
\qty{\sim 45}{\mega\volt\per\meter}~\cite{2009-Padamsee-RFSSTA}.
Recall that,
if the Meissner state of any (type-II) material persists above $\mu_{0}H_{\mathrm{c1}}$,
it must do so in a \emph{metastable} state.
For \gls{dc} fields,
flux penetration can only be prevented by an energy barrier~\cite{1964-Bean-PRL-12-14},
generally anticipated for defect-free surfaces.
From the delineations above,
it is clear that \gls{ltb} results in surfaces that are anything but defect-free
and
any prospect for achieving such high $\mu_{0}H_{\mathrm{vp}}$ fields during
\gls{rf} operation
depends on the interplay between the time needed for the vortex core formation
and
the \gls{rf} period.
In such cases,
maintaining a flux-free state above $\mu_{0}H_\mathrm{c1}$
necessitates the time required for vortex penetration to exceed
the operating \gls{rf} period
(i.e., the inverse \gls{rf} frequency)
of the cavity~\cite{2023-Gurevich-SST-36-063002}.
Comparing this study with $\mu_{0}H_{\mathrm{vp}} \sim \qty{190}{\milli\tesla}$~\cite{2011-Dhavale-AIPCP-1352-119,2012-Dhavale-SST-25-065014},
we suggest that \gls{ltb} cavities need a longer flux nucleation time than the \gls{rf} period to sustain the Meissner state.
Alternatively,
\gls{ss} bilayers can maintain that up to its $\mu_{0}H_{\mathrm{vp}} \sim \qty{235}{\milli\tesla}$,
even in the \gls{dc} flux penetration case,
offering a more robust approach for achieving higher accelerating gradients than \gls{ltb}.

Considering the above details,
it is apparent that \gls{ltb} Nb differs fundamentally from that of an \gls{ss} bilayer
in both its composition and mechanism for impeding flux entry.
Concerning the latter,
our present findings,
in conjunction with related work~\cite{2018-Junginger-PRAB-21-032002,2022-Turner-SR-12-5522},
support the notion that \gls{ltb} does not create a supplemental energy barrier,
but instead postpones vortex penetration above $\mu_{0}H_\mathrm{c1}$ due to pinning.
While it remains an open question as to if this behavior could be further engineered to
benefit \gls{srf} applications,
it is apparent that careful control over the near-surface doping is crucial.
Advances in this area are already apparent~\cite{2021-Lechner-APL-119-082601}.
In future work,
it would be interesting to test these ideas on related
\gls{ltb} treatments~\cite{2013-Grassellino-SST-26-102001,2017-Grassellino-SST-30-094004,arXiv:1806.09824},
as well as the recently discovered
``mid-$T$'' treatments that are known to produce very small
surface resistances~\cite{2020-Posen-PRA-13-014024,2021-Ito-PTEP-2021-071G01}.

\section{
  Conclusion
  \label{sec:conclusion}
 }

Using \gls{musr},
we measured the depth-dependence
(on the \unit{\micro\meter} scale)
of the vortex penetration field in Nb ellipsoids
that received either a \gls{ltb} surface-treatment
or
a \qty{2}{\micro\meter} coating of Nb$_3$Sn.
In each sample,
the measured field of first-flux-entry is greater than Nb's lower critical field of \qty{\sim 170}{\milli\tesla},
suggesting their applicability for \gls{srf} cavities.
In the coated sample,
we find a depth-independent $\mu_{0} H_{\mathrm{vp}} = \qty{234.5 \pm 3.5}{\milli\tesla}$,
consistent with Nb's superheating field and the presence of interface energy barrier
preventing flux penetration.
Conversely,
in \gls{ltb} Nb,
its $\mu_{0} H_{\mathrm{vp}}$ is only moderately larger than Nb's $\mu_{0} H_\mathrm{c1}$,
increasing slightly with increasing depths below the surface.
The latter observation,
in conjunction with the increased span of the Meissner-vortex transition,
suggests pinning from surface-localized defects.
Our findings confirm that the introduction of a thin superconducting overlayer on Nb
can effectively push the onset of vortex penetration up the superheating field,
but rules our \gls{ltb} as a means of achieving this.
We suggest that its success is rather due to effects specific to the operation under \gls{rf} fields,
such as the time required for vortex nucleation.  These findings validate the potential of employing superconducting bilayers to achieve a flux-free Meissner state up to the superheating field of the substrate.

\begin{acknowledgments}
  We thank D.~L.~Hall (Cornell)
  and
  B.~Waraich (TRIUMF)
  for providing the coated and surface-treated Nb samples,
  respectively.
  Technical support during the \gls{musr} experiments from
  R.~Abasalti, D.~J.~Arseneau, B.~Hitti, G.~D.~Morris, and D.~Vyas
  (TRIUMF) is gratefully acknowledged.
  Financial support was provided by the \gls{nserc}
  [SAPPJ-2020-00030 and SAPIN-2021-00032].
\end{acknowledgments}

\section*{author declerations}
	 \subsection*{Conflict of Interest}
	 The authors have no conflicts to disclose.

\section*{data availability}
Raw data of \gls{musr} experiments are publicly available for download from: \href{http://musr.ca/}{http://musr.ca}.

\bibliography{references.bib}

\end{document}